# Compact frequency multiplexed readout of silicon quantum dots in monolithic FDSOI 28nm technology

Quentin Schmidt[1], Baptiste Jadot[1], Brian Martinez[1], Thomas Houriez[1], Adrien Morel[6], Tristan Meunier[3,4], Gaël Pillonnet[1], Gérard Billiot[1], Aloysius Jansen[2], Xavier Jehl[2], Yvain Thonnart[5], Franck Badets[1]

[1]Univ. Grenoble Alpes, CEA, Leti, [2]Univ. Grenoble Alpes, CEA, IRIG, Pheliqs, [3]Univ. Grenoble Alpes, CNRS, Institut Néel, [4]Quobly, [5]Univ. Grenoble Alpes, CEA, List, F-38000 Grenoble, [6]Univ. Savoie Mont Blanc, SYMME, F-74000 Annecy, France
E-mail : quentin.schmidt@cea.fr

*Abstract*—This paper demonstrates the first on-chip frequency multiplexed readout of two co-integrated single-electron transistors without the need for bulky resonators. We characterize single electron dynamics in both single electron transistors at 4.2K before validating their simultaneous readout within 2.2µs, achieving a 99.9% fidelity with a 1MHz frequency spacing. This experimental demonstration paves the way towards resonator-free large-scale quantum-classical architectures that would be required for future universal and reprogrammable quantum computers.

## I. Introduction

Quantum computers will require hundreds of thousands of physical qubits that will need to be biased, addressed, and read. As for readout, silicon spin qubits must achieve >99.9% fidelity combined with a µs readout time ($T_i$) while maintaining the total power consumption below 1W to remain compatible with cryogenic (<1K) operation. Furthermore, for quantum error correction, fault-tolerant qubit networks preferably need to be read simultaneously [1]. The integration of cryogenic circuits (Cryo-CMOS) appears as an effective answer for large-scale multiplexed readout, as it could address the restricted wiring capability of cryostats [1-5]. A few parallel readout architectures based on reflectometry have already been proposed in literature, but they require one bulky inductor-based resonator per qubit, making them hard to scale up to the future required number of qubits, Fig. 1 [1-5]. An alternative approach consists in using single electron transistors (SETs) in order to convey state information into nA-range current, followed by an amplification stage made of a single cryogenic transimpedance amplifier (TIA). However, previous implementations [6, 7] do not meet the bandwidth requirements for fast and multiplexed readout, limiting their use to an individual device readout. In [8, 9], integrator-based architectures have been introduced as an alternative achieving fast readout. However, these solutions do not enable the parallel readout of multiple quantum devices and only allow a sequential time multiplexing technique. Recently, [10] proposed a wide bandwidth capacitive feedback TIA (CTIA) operating at 4.2K. Leveraging on this CTIA, this paper demonstrates the first simultaneous measurement of two co-integrated FDSOI SETs using frequency multiplexing without additional bulky resonator, achieving a fidelity of 99.9% for a 2.2µs integration time.

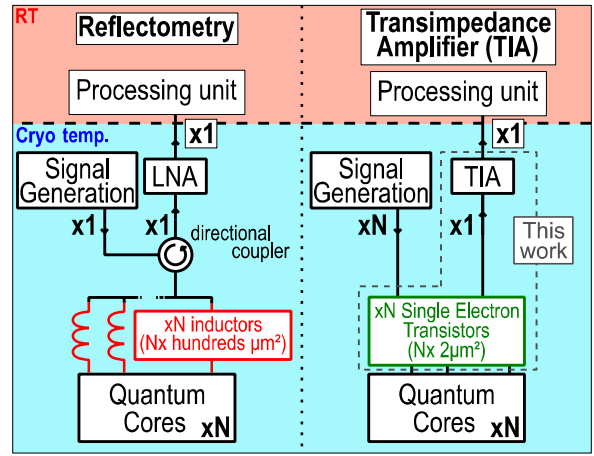

Fig. 1. Block diagram of an envisioned large-scale reflectometry readout compared to the proposed architecture.

## II. Multiplexed charge readout architecture

In the envisioned implementation shown in Fig. 1, each qubit is capacitively coupled to a SET, such that a change in the qubit's spin will induce shifts in both SET's transconductance $g_m$ and output conductance $g_{ds}$. Compared to reflectometry, which relies on inductors, by only using SETs, whose size is comparable to the quantum core itself, the system footprint is minimized, as illustrated in Fig. 1. By modulating respectively the gate or the source with the help of a carrier signal, the $g_m$ or $g_{ds}$ variations could be sensed through the SETs output current amplitude by the readout circuit at the carrier frequency [11]. The proposed measurement setup is illustrated in Fig. 2 and consists in SETs, co-integrated and followed by a single CTIA and a buffer designed in ST28nm FDSOI technology. The DC biases and frequencies generation are brought from room temperature (RT) and the buffer output signal is sent to RT to an on-the-shelf I/Q demodulator. The SETs are made from NMOS FDSOI dual-gate devices with shorted top-gates. FDSOI technology allows high voltage back-biasing ($V_{bg}$=4V) that helps in forming a single dot between both top-gates [12].

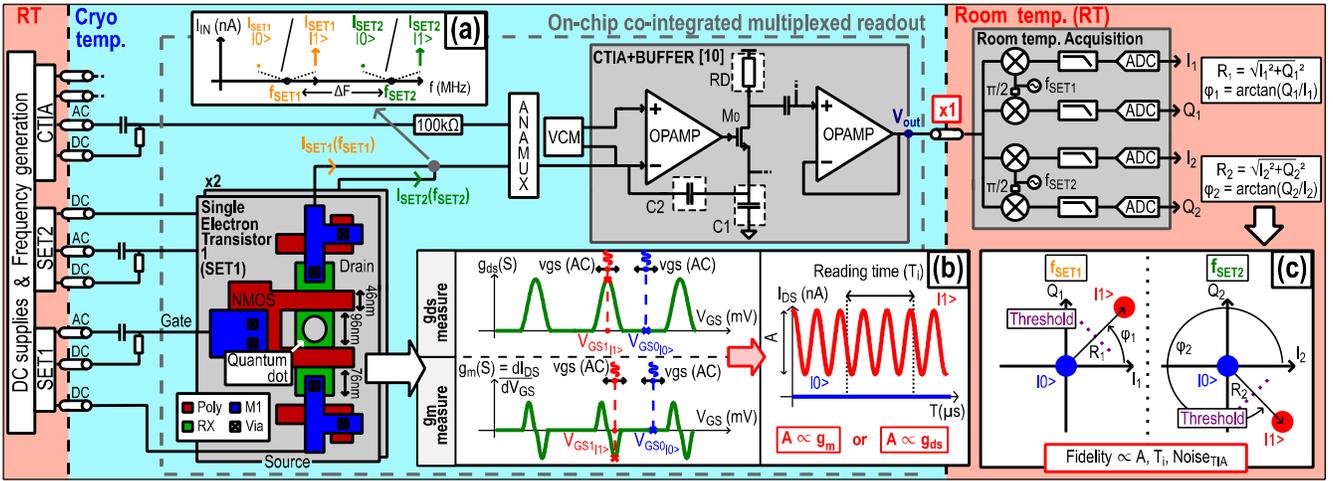

Fig. 2. Single electron transistors (SETs) geometry and block diagram of the proposed frequency multiplexed TIA-based readout. (a) Input signal frequency representation. (b) Illustration of the different measurable parameters. (c) Depiction of the thresholding algorithm for I/Q signals.

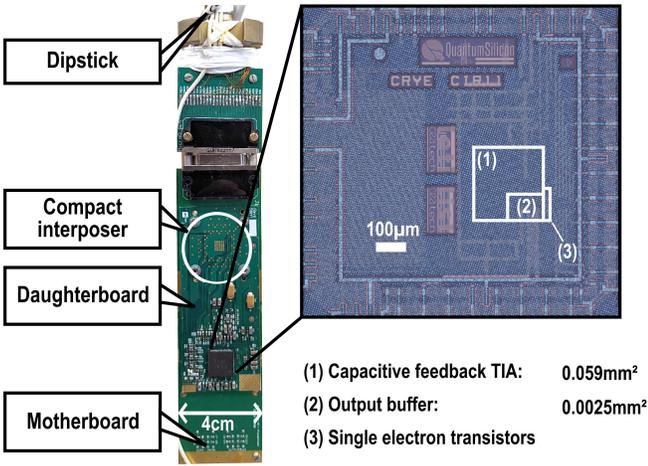

(1) Capacitive feedback TIA: 0.059mm²
(2) Output buffer: 0.0025mm²
(3) Single electron transistors

Fig. 3. Picture of PCB and die micrograph.

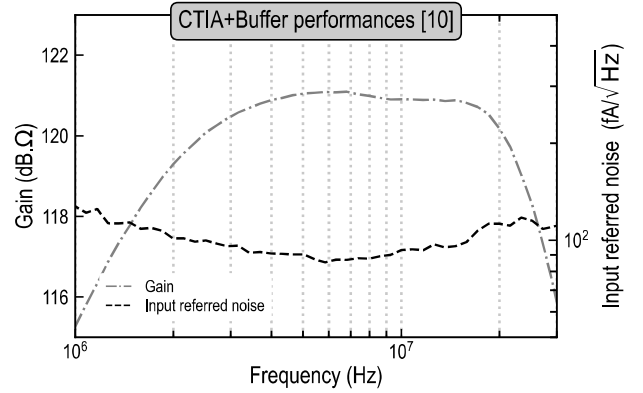

Fig. 4. Standalone cryogenic gain and noise of the proposed amplification chain.

As illustrated in Fig. 2, the proposed multiplexed readout is based on modulating the voltage of either SET's top-gate or source at distinct frequencies in order to extract respectively the $g_m$ or $g_{ds}$ information of both SETs at the same time. The currents drawn from each SET ($I_{DS}$) are summed and amplified by the cryogenic CTIA, then sent to RT via the co-integrated buffer through a single wire. At RT, the signal that contains information on both SETs is demodulated at each frequency using parallel I/Q demodulation schemes and digitized to recover the $g_m$ or $g_{ds}$ of each SET. We define the signal information as the $g_m$ or $g_{ds}$ amplitude difference between states |1⟩ and |0⟩. As shown in Fig. 2b, state |0⟩ is taken arbitrary as the noise floor, i.e. where no DC current pass through the SET (Coulomb blockade), while the SET is tuned so that the state charge of |1⟩ creates a local maximum in $g_m$ or $g_{ds}$, i.e. where a DC current $I_{DS}$ is flowing through the SET. As illustrated in Fig. 2c, the resulting IQ constellations consist in two amplitude levels that are similar to On-Off Keying (OOK) modulation. Furthermore, the frequency multiplexing technique is alike frequency division multiple access (FDMA), where multiple channels are distributed within the system bandwidth. A digital threshold based on the closest mean value is then used to discriminate the parallel measured states, resulting in the readout fidelity and bit error rate (BER).

III. MEASUREMENT RESULTS

Figure 3 shows a micrograph of our 1x1mm² chip, along with the placement of the CTIA, buffer, and SETs. The mother and daughter PCB boards offer flexibility for various testing setups within a compact layout that is constrained by the cryostat.

The standalone cryogenic performance of the CTIA has been first characterized through a 100kΩ co-integrated input resistor that emulates an on-chip current source. As it could be seen in Fig. 4, a 121dB.Ω gain, 25MHz bandwidth and an input referred noise of 90fA/√Hz have been measured for a 428µW power consumption. The analog multiplexer (ANAMUX) illustrated in Fig. 2 allows the characterization of the two co-integrated SETs with the help of the CTIA.

Figure 5 illustrates simultaneous measurement of $g_m$ for both SETs as a function of $V_{gs}$ and $V_{ds}$, using the proposed gate modulation scheme. Each plot illustrates positive/negative signal-patterns that are consistent with the derivative of Coulomb diamond with respect to the gate voltage, as illustrated in Fig. 2b. The asymmetry observed between the two diagrams, e.g. the shape of the Coulomb diamonds as well as the $V_{gs}$ range where they are located, are attributed to the geometrical variability between quantum dots. Additionally, the characteristic of SET2 appear particularly sensitive to at least one two-level fluctuator from its immediate environment, as evidenced by the noise at the black square in Fig. 5b [13]. This validates the ability of SETs to sense their electrostatic environment, and thus offers promising perspectives for future qubit state measurements [11].

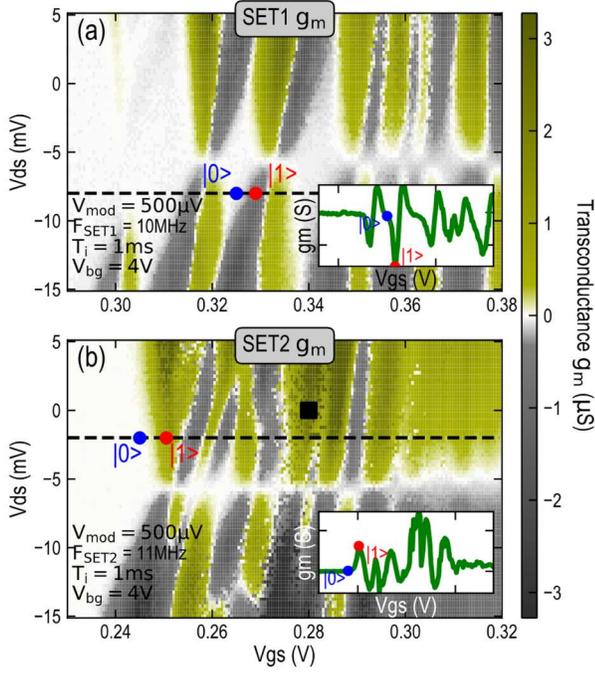

Fig. 5. Frequency multiplexed transconductance ($g_m$) maps for (a) the first and (b) the second SET. The insets of the data traces along the dashed lines illustrate the blue and red bias points for the qubit state readout.

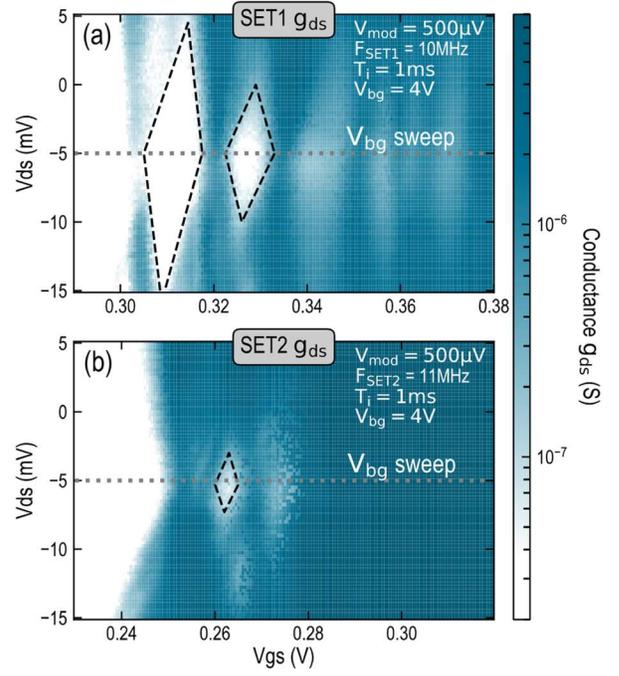

Fig. 6. Frequency multiplexed conductance ($g_{ds}$) maps for (a) the first and (b) the second SET.

The architecture allows another approach based on source modulation and $g_{ds}$ measurement that is illustrated in Fig. 6. As depicted in Fig. 5 and Fig. 6, both measurement techniques lead to the same Coulomb diamond characteristics. As both approaches could present an architectural advantage depending on the selected quantum core network architecture, it confirms the versatility of the proposed readout scheme. In Fig. 7, the conductance $g_{ds}$ is plotted as a function of $V_{gs}$ and $V_{bg}$ for $V_{ds}$=-5mV. A constant slope in the entire characteristics is observed for both SETs as a function of the common back-gate voltage $V_{bg}$. We measured a coupling ratio of approximatively 8.5 between the top-gate and the back-gate, which is in the range of typical characteristic values of forward back-biased ST28nm FDSOI NMOS devices, considering the presence of potential impurities [14]. Therefore, it confirms that the observed characteristics are effectively due to the measured quantum devices.

We emulated qubits states measurement with our two SETs by defining specific $V_{gs}$ and $V_{ds}$ bias couples, as illustrated in Fig. 5 by the red and blue dots in both SETs $g_m$ maps. Using these emulated states, Fig. 8 details the fidelity achieved during a simultaneous 2-channel readout as a function of the integration time $T_i$. For both SETs, we obtain a 99.9% fidelity (BER=$10^{-3}$) in $T_i$=2.2µs, compatible with the aforementioned spin qubit readout requirements. With a frequency separation between channels of ΔF=1MHz, no channel overlap was observed, as predicted by [10], considering that we are well further the theoretical orthogonally frequency-division multiplexing (OFDM) limit in term of frequency separation ($\Delta F > 1/T_i$).

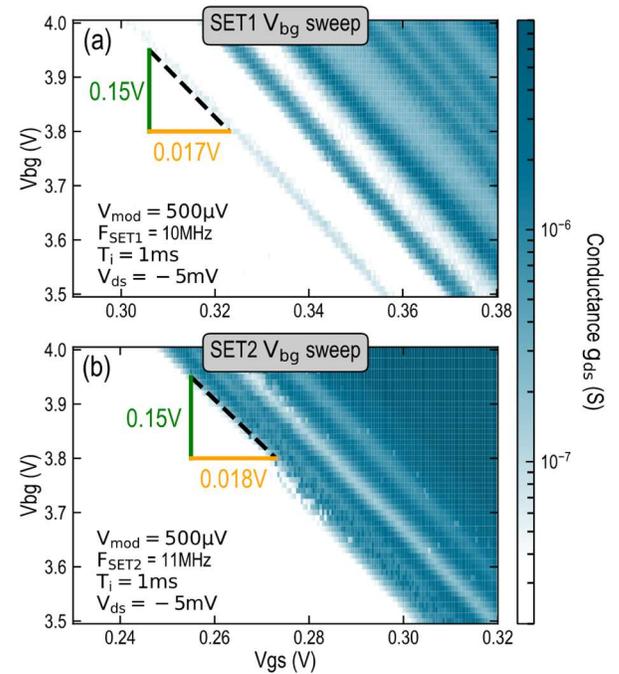

Fig. 7. Frequency multiplexed conductance ($g_{ds}$) maps as a function of the back-gate voltage $V_{bg}$ and $V_{gs}$ for (a) the first and (b) the second SET.

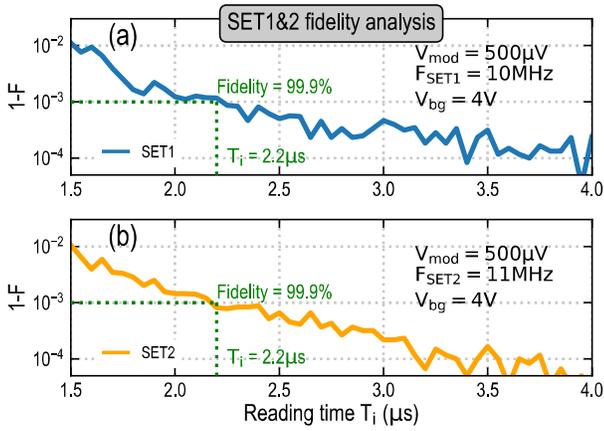

Fig. 8. SETs fidelity (F) as a function of $T_i$ achieved for a simultaneous gate modulation of both SETs.

Table I. presents the comparison of our experimental realization with the existing state-of-the-art of cryo-CMOS readout architectures that were demonstrated with quantum devices. Previous studies based on charge readout [7] have already presented fully integrated architectures. However, the achieved bandwidths were not compatible with the readout time requirements of spin qubits and no parallel readout solution was proposed. Therefore, using charge readout techniques only one QD readout at a time has been demonstrated so far. In [1] a frequency-multiplexed reflectometry-based readout of two QDs, coupled with a time-multiplexing technique, was proposed and achieved the requirements for spin qubits readout. However, the use of bulky inductors raises issues for future large-scale architectures, particularly in terms of space occupation within the cryostat.

## IV. CONCLUSION

This paper presents a parallel measurement of two SETs at 4.2K, highlighting typical Coulomb diamond characteristics. A frequency spacing of 1MHz with no channel overlap has been demonstrated. The sensing of $g_m$ and $g_{ds}$ were validated by the extracted coupling ratio of 8.5 between the top-gate and back-gate, consistent with the typical value for NMOS FDSOI transistors. Both characteristics exhibit expected asymmetry attributed to geometrical variabilities during fabrication. Additionally, the results confirm a particular sensitivity to the nearby electrostatic environment, which is consistent with the charge measurement capabilities of SETs. Finally, this work demonstrate the first fully integrated frequency multiplexed readout of several quantum devices that requires no additional resonators. Moreover, the achieved reading time of 2.2µs with a fidelity of 99.9% meets the requirements for spin qubit readout. This constitutes a first step toward large-scale compatible TIA-based readout architectures that could be integrated into fault-tolerant quantum computers.

TABLE I. EXPERIMENTAL REALIZATION OF QUANTUM DEVICES READOUT WITH CRYO-CMOS

|  | ISSCC20 [7] | Nature22 [1] | This work | Unit |
|---|---|---|---|---|
| **Reading type** | Charge readout | Reflectometry | Charge readout | - |
| **Architecture** | Direct I/Q | Q-C SM[b] | Direct I/Q | - |
| **IC Temperature** | 0.11 | 0.05 | **4.2** | K |
| **Technology** | FDSOI 28nm | Bulk 40nm | **FDSOI 28nm** | - |
| **Off-chip components at cryogenic temp.** | No | Dir. Coupler, LNA | **No** | - |
| **Demonstrated Multiplexing** | No | Time and frequency | **Frequency** | - |
| **Simultaneously evaluated quantum devices** | 1 QD[d] | 2 QDs[d] | **2 QDs[d]** | - |
| **Footprint** | 0.01 | 0.44[a,c] | **0.06** | mm² |
| **Power consumption** | 0.001 | - | **0.428** | mW |
| **Operating frequency** | 0.001 | 6,000-8,000 | **1.5-25** | MHz |

[a] Reported data that doesn't consider discrete cryogenic electronics.
[b] Quantum-classical Switch Matrix.   [c] Estimated from chip micrograph.   [d] Quantum Dot.

## ACKNOWLEDGMENTS

This work is supported by the French National Research Agency under the program "France 2030" (PEPR PRESQUILE - ANR-22-PETQ-0002). Electronic PCBs were supplied by CEA test engineers as part of the European Qu-test project.

## REFERENCES

[1] Ruffino *et al.* "A cryo-CMOS chip that integrates silicon quantum dots and multiplexed dispersive readout electronics", Nat Electron 5, 2022.

[2] Y. Peng *et al.*, "A cryo-CMOS wideband quadrature receiver with frequency synthesizer for scalable multiplexed readout of silicon spin qubits," JSSC, vol. 57, 2022.

[3] J. Park *et al.*, "A fully integrated cryo-CMOS SoC for state manipulation, readout, and high-Speed gate pulsing of spin qubits," JSSC, vol. 56, 2021.

[4] B. Prabowo *et al.*, "13.3 A 6-to-8GHz 0.17mW/qubit cryo-CMOS receiver for multiple spin qubit readout in 40nm CMOS technology," ISSCC, 2021.

[5] A. Nagulu *et al.*, "Sub-mW/qubit 5.2-7.2GHz 65nm cryo-CMOS RX for scalable quantum computing applications," CICC, 2023.

[6] M. L. V. Tagliaferri *et al.*, "Modular printed circuit boards for broadband characterization of nanoelectronic quantum devices," TIM, vol. 65. 2016.

[7] L. L. Guevel *et al.*, "19.2 A 110mK 295µW 28nm FDSOI CMOS quantum integrated circuit with a 2.8GHz excitation and nA current sensing of an on-chip double quantum dot," ISSCC, 2020.

[8] H. Fuketa *et al.*, "A cryogenic CMOS current comparator for spin qubit readout achieving fast readout time and high current resolution," VLSI Technology and Circuits, 2022.

[9] M. Castriotta *et al.*, "Fully integrated cryo-CMOS spin-to-digital readout for semiconductor qubits," SSCL, vol. 6, 2023.

[10] Q. Schmidt *et al.*, " A 7.4µW and 860µm² per channel cryo-CMOS IC for 70-channel frequency-multiplexed µs-readout of semiconductor qubits," CICC, 2024.

[11] Yang *et al.*, "Operation of a silicon quantum processor unit cell above one kelvin", Nature 580, 2020.

[12] L. Le Guevel *et al.*, "Low-power transimpedance amplifier for cryogenic integration with quantum devices", Applied Phy. Rev., vol. 7, 2020.

[13] C. Spence *et al.*, "Probing low-frequency charge noise in few-electron CMOS quantum dots", Phys. Rev. Appl., vol. 19, 2023.

[14] L. Le Guevel et al., « Impedancemetry of multiplexed quantum devices using an on-chip cryogenic complementary metal-oxide-semiconductor active inductor », Chip, vol. 2, 2023